# SpindleFlexNet: Flexible sleep spindles detection for EEG signals based on an adaptive one-dimensional RetinaNet-based framework

Shao-Jun Xia, *Graduate Member, IEEE*, Jing Bao, *Graduate Member, IEEE*, Anlan Sun, Xiaoyang Chen, Hongjia Liu, *Graduate Member, IEEE*, Xiao-Ting Li, and Ying-Shi Sun

***Abstract*—Sleep spindle is a physiologically significant biomedical signal in electroencephalographic (EEG) waveforms, which is typically a low-amplitude event in sleep. Due to the small signal ratio in the overall EEG, previous detection methods have limited capability to capture its start and end points and lack flexibility in handling multi-spindle scenarios. To address the gap, we address the problem from a new perspective and introduce SpindleFlexNet, the first framework in this field to apply deep learning-based one-dimensional object detection, leveraging an adapted one-dimensional RetinaNet architecture. The framework employs one-dimensional anchor generation, matching, and regression, along with a customized one-dimensional loss function. Analyses were conducted on two public datasets: the Montreal Archive of Sleep Studies and DREAMS, from which a total of 11,061 and 335 segments were obtained, respectively. When trained on these datasets, SpindleFlexNet achieved an average recall, precision, and F1-score of 0.61, 0.76, 0.67, and 0.58, 0.80, 0.67 in five-fold cross-validation. The model demonstrates stable detection performance and good generalization, making it a practical tool for sleep research. Potential applications include automated spindle labeling in clinical settings and as a reference for studies combining EEG with simultaneous functional magnetic resonance imaging.**

## I. Introduction

Sleep spindles are transient bursts of oscillatory brain activity observed in electroencephalography (EEG), primarily during non-rapid eye movement sleep stage [1, 2]. Sleep spindles typically last 0.5–1.5 seconds, occur within the 11–16 Hz frequency range, and exhibit amplitudes between 20–40 μV [3-5]. Due to the temporal and spectral characteristics, sleep spindles are widely regarded as reliable neurophysiological markers in sleep research and cognitive neuroscience [6]. Accurate detection of sleep spindles is crucial for understanding sleep states and assessing neural function [7, 8]. Numerous studies have associated spindle activity with higher-order cognitive processes such as sensory gating and memory consolidation [9-11]. Interest in spindle analysis is further driven by observed alterations in spindle features across various neurological and psychiatric conditions, such as epilepsy, neurodegenerative diseases, and mood disorders [12-16], highlighting the value of robust and automated sleep spindles detection algorithms in both clinical diagnostics and large-scale sleep research applications.

Expert visual annotations remain the gold standard for sleep spindles detection [17]. In the absence of automated tools, spindle events are manually annotated on EEG recordings by trained sleep specialists. However, this process is inherently subjective, often leading to low inter-rater agreement due to differences in expert experience [18, 19]. For example, as widely used open datasets, Montreal archive of sleep studies database [20] (MASS, http://www.ceams-carsm.ca/mass) and DREAMS database (DREAMS, https://zenodo.org/record/2650142#.Yu_n6XpBxPY) both have two significantly varied annotations from two different sleep experts. Furthermore, sleep spindles constitute only a small fraction of the overall EEG signal, making manual annotation both time-consuming and labor-intensive [21]. These limitations significantly hinder the scalability and clinical utility of spindle analysis, thereby highlighting the pressing need for reliable and efficient automated detection methods.

Early automated detection methods for sleep spindles primarily rely on heuristic rules such as amplitude thresholding [22, 23], band-pass filtering [24], band-pass energy [25, 26], short-time Fourier transform [27], wavelet transform [28, 29], etc. Subsequent studies [30-33] improved

*(Corresponding authors: Anlan Sun, Ying-Shi Sun.)* Shao-Jun Xia, Jing Bao, and Anlan Sun contributed equally to this work and are co-first authors.

Shao-Jun Xia, Xiao-Ting Li, and Ying-Shi Sun are with the Key Laboratory of Carcinogenesis and Translational Research (Ministry of Education/Beijing), Department of Radiology, Peking University Cancer Hospital & Institute, Beijing 100142, China (e-mail: xsjacademic@163.com; 13520120308@163.com; sys27@bjmu.edu.cn).

Jing Bao is with the Department of Signal and Image Processing, Centrale Nantes, Nantes 44300, France (e-mail: BaoJing@ieee.org).

Anlan Sun is with Yizhun Medical AI Co., Ltd., Beijing 100191, China (e-mail: anlan.sun@yizhun-ai.com).

Xiaoyang Chen is with the Department of Radiology, University of North Carolina at Chapel Hill, Chapel Hill, NC 27599, USA, and Biomedical Research Imaging Center, University of North Carolina at Chapel Hill, Chapel Hill, NC 27599, USA (e-mail: Xiaoyang_chen@med.unc.edu).

Hongjia Liu is with the Department of Radiation Oncology, Peking University Shenzhen Hospital, Shenzhen 518033, China (e-mail: stevendeliu@foxmail.com).

these approaches by incorporating additional signal features, including root mean square, signal envelope, short-time energy, and zero-crossing rate, applied to σ-band (11–16 Hz) filtered signals, which led to moderate improvements in recall. Further developments [34] integrated multiple σ-band power and correlation-based features within fixed temporal windows (e.g., 0.3–2.5 s), achieving better precision and recall. More recent refinements [35-37] introduced techniques such as background normalization, instantaneous bandwidth estimation, and coupling with slow-wave activity, although these enhancements resulted in marginal performance gains. Overall, these methods heavily depend on manually designed features and threshold tuning, which limits their generalizability across datasets due to variability in empirical parameter settings.

Machine learning techniques have also been applied for sleep spindle detection. For example, tree-based ensembles, support vector machines [19], boosting [38], bagging [39], and probabilistic classifiers [40] have shown robust performance for spindle detection across diverse EEG datasets. However, these approaches typically have two stages, relying on traditional methods for feature extraction. Moreover, most are primarily classification-oriented and lack the flexibility to precisely determine the onset and offset of sleep spindles.

To address these limitations, recent studies have explored the use of deep learning for spindle detection, e.g. SpindleNet [41], transfer learning classifiers [42], and U-Net architectures [43, 44] , neither relying on empirical nor constrained features by prior expertise, while enabling end-to-end and flexible detection. Kulkarni et al. [41] proposed SpindleNet to detect sleep spindles based on a single EEG channel. While the model is effective for real-time detection and provides high classification performance in identifying spindles, its architecture cannot directly perform coordinate regression and relies on multiple complex subnetworks. Transfer learning classifiers [42] can effectively leverage pre-trained weights and also achieve high classification performance, yet they are inherently limited to short EEG segments and cannot provide coordinate-level predictions for spindle onset and offset. U-Net models [43, 44] can provide probability predictions and good temporal resolution, but the U-Net architecture remains relatively heavy, and well performance relies on the subjective design of the additive attention gates. These drawbacks motivate the development of more robust, lightweight, and clinically applicable solutions.

Given the demonstrated capabilities in recognizing and localizing complex patterns, deep learning-based object detection algorithms have achieved significant advances in computer vision, e.g. R-CNN [45], YOLO [46], and RetinaNet [47], and have been adapted for one-dimensional signal analysis, such as electrocardiogram [48] and electromyogram [49]. However, adapting these algorithms for one-dimensional sleep spindle detection has not yet been explored. Directly adapting a 1D object detector using standard object detection algorithms to sleep spindles detection faces two major challenges: (1) Spindles have variable durations (0.5-3 s), variable amplitudes, and occur sparsely in the EEG with large segments of non-spindle activity. Thus, localization and classification require fine temporal resolution and the ability to distinguish large non-spindle negative regions. (2) Standard detection networks often employ deep backbones, which causes the receptive field expand rapidly. An excessively large receptive field implies that the response of an output unit depends on a long temporal window of the input. This process effectively smooths the onset and offset of transient events, leading to imprecise temporal boundaries. In vision or image detection tasks, such blurring may not pose a critical issue. However, for short-duration events with fine spectral characteristics, this boundary blurring can substantially reduce detection accuracy. Therefore, the lack of integration between existing object detection frameworks and one-dimensional EEG-based spindle detection represents a critical gap in the field.

The objective of this study is to develop a reliable and efficient one-dimensional method for automated sleep spindles detection. We present a novel, object detection-inspired deep learning framework, termed SpindleFlexNet, for this purpose. The main contributions of the paper include three aspects: (1) We propose a flexible sleep spindle detection framework by leveraging a one-dimensional adaptation of RetinaNet. Our work represents the first application of a deep learning-based object detection algorithm in the sleep spindle domain. (2) To adapt anchor-based detection for one-dimensional spindle signals, we redesign the entire pipeline into a form tailored for 1D signal representation, enabling robust classification and onset and offset localization. (3) The approach achieves dense anchor-based temporal event detection of multiple sleep spindles (1~7) in 15-second long EEG segments, along with strong temporal sensitivity and high detection efficiency.

## II. Materials and Methods

### A. Data

Two publicly available sleep EEG datasets, MASS and DREAMS, were used in this study. Sleep spindle annotations were obtained directly from the datasets, which provide expert-labeled spindle events. The MASS dataset used in this study corresponds to the SS2 subset. MASS dataset includes EEG recordings from 19 subjects. Among them, 15 subjects were used in our research, which are labeled independently by two experts (denoted as E1 and E2). The EEG signals were acquired using a 32-channel PSG setup, with each subject undergoing approximately 8 hours of recording. All data were sampled at 256 Hz, and C3–A1 channel was picked for analysis. DREAMS dataset consists of 30-minute EEG recordings from 8 subjects. Spindles for six subjects were annotated by two independent experts (denoted as E3 and E4). Annotations were based on the Cz–A1 channel at 200 Hz, except for Subjects 1 and 3, whose signals were obtained from the C3–A1 channel at 100 Hz and 50 Hz, respectively. The signals from the aforementioned channels were likewise resampled to 256 Hz. In total, 21 Participants with dual expert annotations (15 from MASS and 6 from DREAMS) were involved in the final analysis to validate the robustness and generalizability of the proposed detection approach.

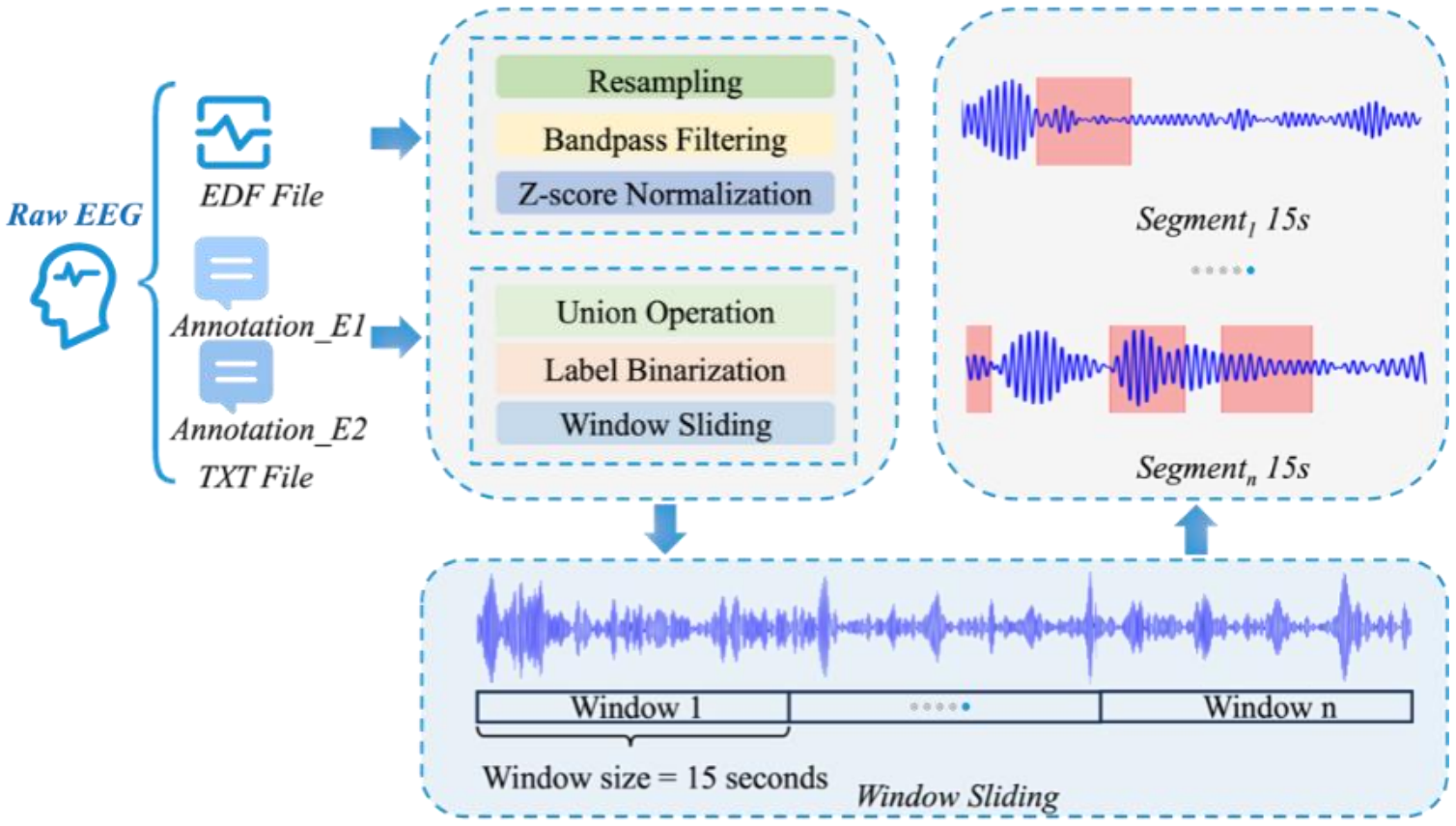

Fig. 1. Flowchart of the preprocessing procedure for spindle segment extraction.

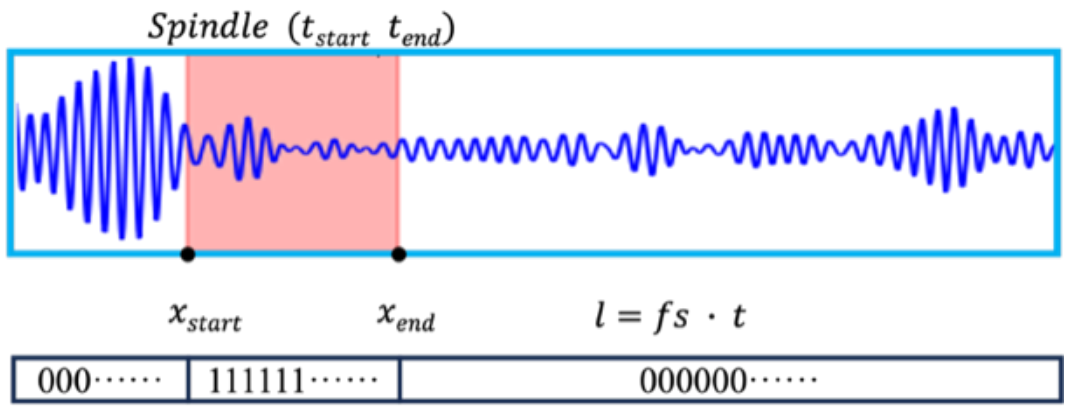

Fig. 2. Illustration of a spindle segment (top) and the corresponding binary label (bottom).

## B. Data Pre-Processing and Segments Extraction

In data pre-processing, the EDF files from MASS and DREAMS datasets were loaded using MNE Python Package (v1.3.1). Signals were resampled to a unified frequency of 256 Hz when necessary to standardize temporal resolution. A sleep spindle is characterized as a transient oscillatory event within the 11–16 Hz frequency range and lasting at least 0.5 s, generally less than 1.5 s, as defined by the American Academy of Sleep Medicine. Accordingly, a bandpass filter (11–16 Hz) was applied to extract spindle-related activity, and events shorter than 0.5 s or longer than 3.0 s were excluded. The filtered signals were subsequently z-score normalized to reduce inter-subject amplitude variability. Band-pass filtering isolates the typical spindle frequency range but does not remove inter-individual spectral variability. Individual slow and fast spindle components remain preserved in the filtered signal.

Spindle annotations were extracted from TXT files independently labeled by two experts. To construct a unified standard, the annotations from two experts were combined via union operation (i.e., E1 ∪ E2, E3 ∪ E4). The normalized EEG signals were segmented by a series of non-overlapping 15-second windows, retaining only those segments that contained at least one spindle event. For each segment, the binary labels of the included spindles were generated to indicate the relative positions within the 15-second interval. All the resulting EEG segments and binary labels were stored in NPY files for classification and regression in our model. The overall data preprocessing pipeline is shown in Figure 1, and a representative EEG signal segment and its corresponding binarized label are illustrated in Figure 2, respectively.

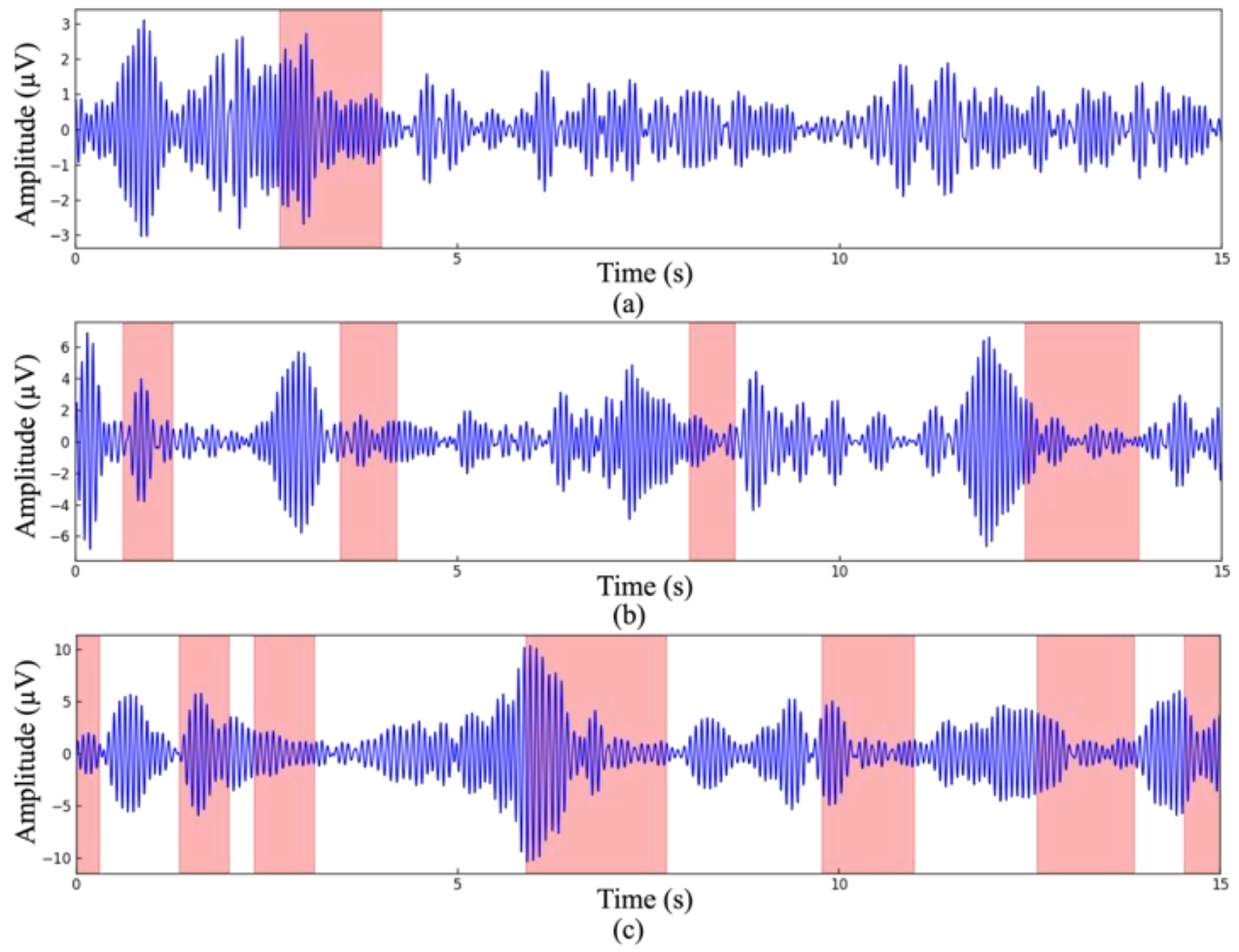

Fig. 3. Examples of three spindle segments (red box) with one, four, and seven spindles, respectively.

## C. SpindleFlexNet Model: Overview of the Architecture

Inspired by the original RetinaNet architecture, we propose a flexible one-dimensional variant, i.e. SpindleFlexNet, for sleep spindled detections. Using one-dimensional convolutions, we adapted the original RetinaNet architecture into a one-dimensional form and designed a set of mechanisms specific to the characteristics of the spindle detection task. Temporal localization of sleep spindle events along one-dimensional EEG signals was fundamentally implemented by employing a one-dimensional anchor mechanism. Anchors are pre-defined temporal intervals placed densely across the input signal, serving as reference candidates for predicting the onset and duration of spindles. A spindle in the EEG signal has onset and offset times denoted by $t_{start}$ and $t_{end}$, respectively, with a duration $t$ and a sampling frequency $fs$. The onset coordinate $x_{start}$ and offset coordinate $x_{end}$ in the segment can be obtained by the product of the time and the sampling frequency. Given the center location $x$ and width $l$ of the ground truth box, each anchor is defined by its onset time $x_{a_{start}}$, offset time $x_{a_{end}}$, center time $x_a$ and temporal length $l_a$, forming a fixed window over the time axis. The one-dimensional regression targets are defined as follows:

$$x = \frac{x_{start} + x_{end}}{2} \tag{1}$$

$$l = fs \cdot t = fs \cdot (t_{end} - t_{start}) = x_{end} - x_{start} \tag{2}$$

$$u_x = \frac{x - x_a}{l_a} \tag{3}$$

$$u_l = \log\left(\frac{l}{l_a}\right) \tag{4}$$

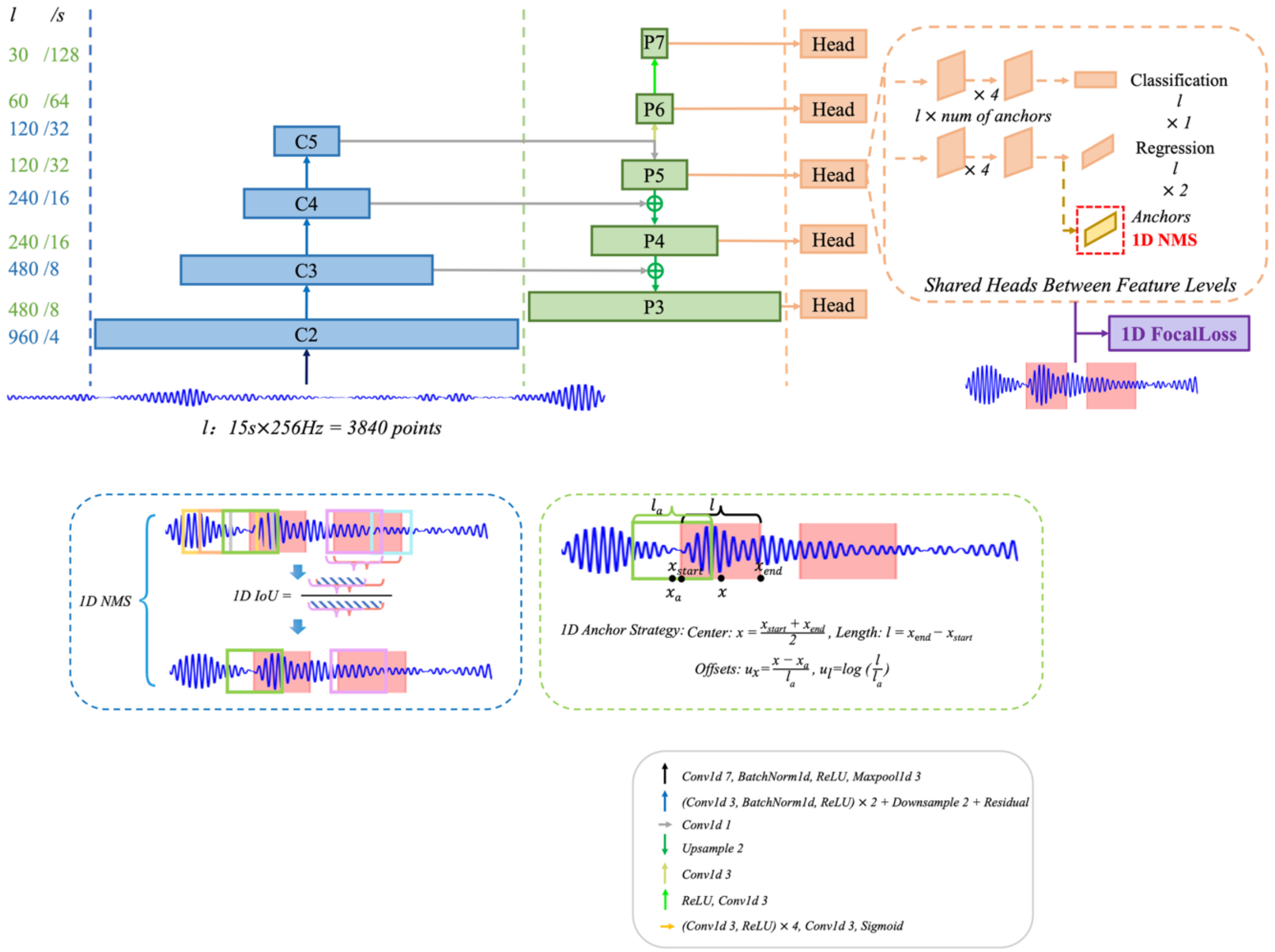

Fig. 4. Overall framework of the proposed SpindleFlexNet model for spindle classification and regression.

Where $u_x$ represents the normalized offset between the ground truth center and the anchor center, $u_l$ denotes the logarithmic ratio of the ground truth width to the anchor width. At each temporal location of the feature map, multiple anchors were generated with duration ratios [0.65, 0.85, 1.0, 1.3, 1.8] and scales [0.85, 1.0, 1.3], similar to the anchor configuration in RetinaNet, to accommodate the variable lengths of spindles. These parameters were selected based on the empirical distribution of spindle lengths observed in training data. So the equation can be written as:

$$w_a = w_0 \cdot s \cdot \sqrt{r} \tag{5}$$

where $w_0$ is the base width, $s$ is the scale factor, and $r$ is the duration ratio.

Building upon the one-dimensional anchor mechanism, the overall framework comprises three key components: a one-dimensional feature extraction backbone, a temporal feature pyramid network (FPN), and two parallel subnetworks for classification and regression. The backbone network is a one-dimensional ResNet18 designed to extract hierarchical temporal features from raw EEG segments. It consists of multiple convolutional layers with progressively increasing receptive fields to capture both short- and long-duration patterns, followed by batch normalization and ReLU activations. The extracted multi-scale features are then passed into a one-dimensional FPN, which fuses temporal features across multiple resolutions using top-down and lateral connections. This design ensures the preservation of fine-grained temporal information essential for accurately detecting spindle onsets and durations. Two specialized subnetworks operate in parallel on the output of the FPN: (1) The classification subnet predicts the presence or absence of spindles within predefined anchor intervals. (2) The regression subnet estimates the temporal onsets and durations relative to the anchor locations, refining the spindle boundaries. The proposed one-dimensional RetinaNet framework allows for end-to-end learning of both localization and classification tasks, making it well-suited for flexible and robust sleep spindle detection in EEG signals.

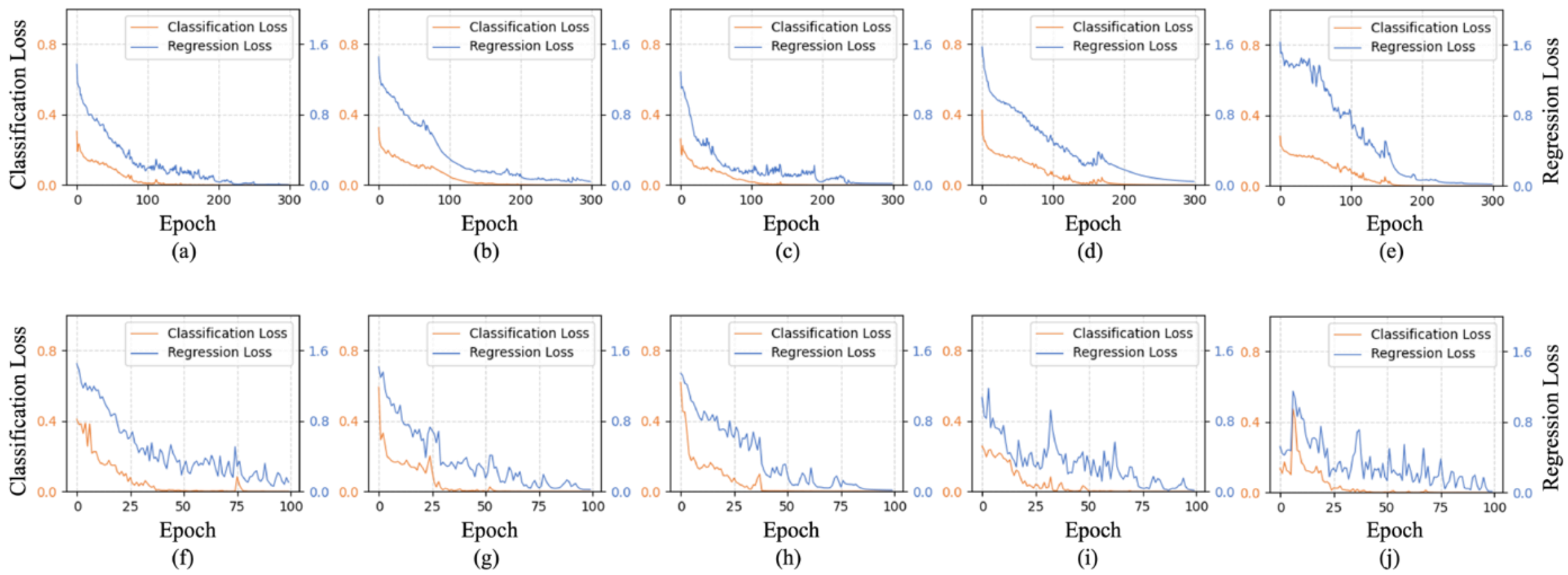

Fig. 5. Learning curves of training via five folds on the MASS and DREAMS datasets.

### D. Modified IoU, Refined Focal Loss, and Total Loss Function

To evaluate the overlap between prediction intervals and ground truth, a one-dimensional IoU metric was introduced for the calculation. Given an anchor interval $P = [x_{a_{start}},\ x_{a_{end}}]$ and the ground truth $G = [x_{start},\ x_{end}]$, the one-dimensional IoU is defined as:

$$IoU\ (P, G) = \frac{\max(0, \min(x_{a_{end}}, x_{end}) - \max(x_{a_{start}}, x_{start}))}{\max(x_{a_{end}}, x_{end}) - \min(x_{a_{start}}, x_{start})} \quad (6)$$

If the IoU between the predicted result and the ground truth exceeds a predefined threshold (i.e., 0.3), the prediction is considered a true positive. This metric is specifically adapted to capture the temporal alignment between predicted and annotated spindle events.

Since sleep spindles are sparse-event scenarios in the overall EEG signals, a modified Focal Loss was used during training to handle the high imbalance between spindle and non-spindle intervals. The Focal Loss $\mathcal{L}_{focal}$ modulates the standard binary cross-entropy loss $CE$ by adding a factor that down-weights easy negatives: where $p_t$ is the model's estimated probability for the true class, $\alpha_t$ is a class-balancing coefficient, and $\gamma$ is the focusing parameter that adjusts the rate at which easy examples are down-weighted (commonly set to 2.0). This formulation effectively emphasizes hard-to-classify spindle events while reducing the influence of abundant non-spindle intervals.

$$CE(p, y) = \begin{cases} -\log(p), & if\ y\ =\ 1 \\ -\log(1-p), & otherwise \end{cases} \quad (7)$$

$$p_t = \begin{cases} p, & if\ y\ =\ 1 \\ 1-p, & otherwise \end{cases} \quad (8)$$

$$CE(p_t) = -\alpha_t log(p_t) \quad (9)$$

$$\mathcal{L}_{focal}(p_t) = -\alpha_t(1-p_t)^{\gamma} log(p_t) \quad (10)$$

According to the above definitions, the total loss function $\mathcal{L}_{total}$ consists of a classification term $\mathcal{L}_{cls}$ based on Focal Loss and a regression term $\mathcal{L}_{reg}$ based on Smooth L1 Loss. In Smooth L1 Loss, $u$ is the prediction error and $\delta$ is the threshold that controls the transition between L1 and L2 Loss.

$$\mathcal{L}_{total} = \frac{1}{N}\sum \mathcal{L}_{cls} + \frac{1}{N_{pos}}\lambda\mathcal{L}_{reg} \quad (11)$$

$$\mathcal{L}_{cls} = \mathcal{L}_{focal}(p_t) \quad (12)$$

$$\mathcal{L}_{reg} = Smooth\ L1(u) \quad (13)$$

$$Smooth\ L1(u) = \begin{cases} 0.5 \cdot \dfrac{u^2}{\delta} & if\ |u| < \delta \\ |u| - 0.5 \cdot\ \delta & otherwise \end{cases} \quad (14)$$

Where $N$ is the total number of samples and $N_{pos}$ represents the number of positive samples.

### E. Post-Processing: One Dimensional NMS

**Algorithm 1** One Dimensional NMS

Boxes $B = \{(x^k_{\text{start}}, x^k_{\text{end}})\}_{k=1}^{n}$, $x^k_{\text{start}} < x^k_{\text{end}}$
scores $s \in \mathbb{R}^n$, threshold $\tau \in [0, 1]$
$\mathcal{K} \leftarrow \varnothing$
**if** $n = 0$ **then**
  **return** $\mathcal{K}$
$O \leftarrow \text{argsort}(s,\ \text{descending})$
**while** $O \neq \varnothing$ **do**
  $i \leftarrow O[1]$
  $\mathcal{K} \leftarrow \mathcal{K} \cup \{i\}$
  $O \leftarrow O \setminus \{i\}$
  **for all** $j \in O$ **do**
    $a_i \leftarrow x^i_{\text{end}} - x^i_{\text{start}}, \quad a_j \leftarrow x^j_{\text{end}} - x^j_{\text{start}}$
    $x_1 \leftarrow \max(x^i_{\text{start}}, x^j_{\text{start}}), \quad x_2 \leftarrow \min(x^i_{\text{end}}, x^j_{\text{end}})$
    $\text{inter}_{ij} \leftarrow \max(0, x_2 - x_1)$
    $\text{IoU}_{ij} \leftarrow \dfrac{\text{inter}_{ij}}{a_i + a_j - \text{inter}_{ij}}$
  $O \leftarrow \{j \in O \mid \text{IoU}_{ij} \leq \tau\}$
**return** $\mathcal{K}$

During inference, the model may produce multiple overlapping detections for the same spindle event. A one-dimensional version NMS is applied to eliminate redundant predictions. Given a list of predicted intervals with associated confidence scores, the 1D NMS iteratively selects the highest-scoring interval and suppresses all other overlapping intervals for which 1D IoU with the selected interval exceeds a suppression threshold (i.e., 0.05). This ensures that only the most confident and non-overlapping detections are retained.

For a series of predicted intervals with corresponding confidence scores, the 1D NMS proceeds as illustrated in the pseudocode diagram: (1) Sort all intervals by score in descending order; (2) Select the interval with the highest score and remove it from the list; (3) Remove all remaining intervals with IoU greater than the threshold with the selected interval; (4) Repeat steps 2–3 until no intervals remain.

### F. Implementation details on training

The union of expert annotations (E1 ∪ E2, E3 ∪ E4) from MASS and DREAMS were obtained by Python 3.9 on PyCharm software (2024 version). The network architecture was developed by PyTorch 1.12.1 (Linux system, Ubuntu 18.04). Deep learning models referring to different folds on the two datasets were trained on a workstation with one GeForce RTX 3080 GPU. The training processes were conducted with batch sizes of 64 on MASS and 4 on DREAMS, with an initial learning rate of 1e-4 using AdamW optimizer. The learning rate was adjusted via (function: ReduceLROnPlateau; factor = 0.5, patience = 5). All training epochs were set to 300. Learning curves of five folds on the MASS and DREAMS datasets are depicted in Figure 5.

### G. Metrics for Performance Evaluation

Average Precision (AP): Average precision is a performance metric that quantifies the area under the precision-recall curve, which summarizes model performance by integrating precision and recall over all confidence thresholds into a single scalar value.

$$AP = \int_0^1 Precision(Recall)\, d(Recall) \qquad (15)$$

Recall: Recall calculates the ratio of true positives to the total positives, i.e. the sum of TPs and false negatives (FN), representing the sensitivity of the model to positive cases.

$$Recall = \frac{TP}{TP + FN} \qquad (16)$$

Precision: Precision essentially measures the proportion of the true positive (TP) samples in all predicted positive samples, including TP and false positive (FP) cases, indicating the accuracy of positive predictions.

$$Precision = \frac{TP}{TP + FP} \qquad (17)$$

F1-score: $F_1$-score serves as a trade-off metric between precision and recall, reflecting the balanced accuracy of the model. Numerically, F1-score is the harmonic mean of precision and recall.

$$F_1 - score = \frac{2Precision \cdot Recall}{Precision + Recall} \qquad (18)$$

.

$$IoU = \frac{P \cap G}{P \cup G} \qquad (19)$$

## III. Results

### A. Characteristics of Extracted Spindle Segments

All extracted segments were 15 seconds in duration, with each varying from one to seven labeled spindles. For anchor-based temporal detection, the 15-second windows accounts for the sparse, short, and randomly occurring nature of spindle events in whole-night EEG, while ensuring effective anchor–target matching within a controlled temporal context. Due to the sampling rate of 256 Hz, each segment contains 3840 time points. Similarly, the time points of each spindle within a segment are equal to the product of its duration and 256 Hz.

TABLE I
PARTICIPANT INFORMATION FOR THE MASS DATASETS

| Subject ID | No. of segments | Max spindles | Minimum spindle period (s) | Maximum spindle period (s) |
|---|---|---|---|---|
| MASS | | | | |
| 1 | 898 | 7 | 0.50 | 2.94 |
| 2 | 1024 | 7 | 0.50 | 2.97 |
| 3 | 361 | 5 | 0.50 | 2.67 |
| 5 | 594 | 6 | 0.50 | 2.52 |
| 6 | 531 | 4 | 0.50 | 3.00 |
| 7 | 765 | 7 | 0.50 | 3.00 |
| 9 | 767 | 6 | 0.50 | 2.96 |
| 10 | 866 | 6 | 0.50 | 2.91 |
| 11 | 720 | 6 | 0.52 | 3.00 |
| 12 | 661 | 6 | 0.52 | 2.96 |
| 13 | 856 | 5 | 0.50 | 3.00 |
| 14 | 782 | 7 | 0.50 | 3.00 |
| 17 | 728 | 5 | 0.50 | 2.97 |
| 18 | 843 | 7 | 0.50 | 2.91 |
| 19 | 666 | 5 | 0.50 | 2.79 |

TABLE II
PARTICIPANT INFORMATION FOR THE DREAMS DATASETS

| Subject ID | No. of segments | Max spindles | Minimum spindle period (s) | Maximum spindle period (s) |
|---|---|---|---|---|
| DREAMS | | | | |
| 1 | 79 | 5 | 0.50 | 1.72 |
| 2 | 53 | 4 | 0.51 | 1.46 |
| 3 | 35 | 3 | 1.00 | 1.63 |
| 4 | 45 | 4 | 0.50 | 1.80 |
| 5 | 58 | 4 | 0.50 | 1.28 |
| 6 | 65 | 5 | 0.50 | 1.39 |

Correspondingly, the labels were also converted into point-based binary masks to fit the input of SpindleFlexNet, as shown in Figure 2. Examples of segments are demonstrated in Figure 3. Each segment contains one, four, and seven spindles highlighted by red boxes, respectively. Table I and Table II summarize the number of 15-second segments extracted from each participant on the MASS and DREAMS datasets, along with typical characteristics such as segment-wise maximum number of spindles, minimum spindle period, and maximum spindle period. In total, 11,061 segments were obtained from MASS and 335 segments from DREAMS.

### B. Cross-Validation Performance on Benchmark Datasets

TABLE III
EVALUATION METRICS OF THE MODEL UNDER FIVE-FOLD CROSS-VALIDATION ON THE MASS AND DREAMS DATASETS, USING THE UNION ANNOTATIONS FROM TWO EXPERTS.

| Subject ID | Validation fold | AP | Recall | Precision | F1-score | IoU |
|---|---|---|---|---|---|---|
| | | | MASS | | | |
| 1, 2, 3 | 1 | 0.58 | 0.56 | 0.76 | 0.64 | 0.43 |
| 5, 6, 7 | 2 | 0.61 | 0.56 | 0.70 | 0.62 | 0.24 |
| 9, 10, 11 | 3 | 0.60 | 0.60 | 0.69 | 0.64 | 0.26 |
| 12, 13, 14 | 4 | 0.79 | 0.66 | 0.83 | 0.73 | 0.25 |
| 17, 18, 19 | 5 | 0.78 | 0.68 | 0.80 | 0.73 | 0.24 |
| Mean | | 0.67 | 0.61 | 0.76 | 0.67 | 0.28 |
| | | | DREAMS | | | |
| 1 | 1 | 0.72 | 0.61 | 0.81 | 0.69 | 0.33 |
| 2 | 2 | 0.75 | 0.60 | 0.86 | 0.71 | 0.28 |
| 3, 4 | 3 | 0.42 | 0.40 | 0.61 | 0.48 | 0.21 |
| 5 | 4 | 0.72 | 0.66 | 0.85 | 0.74 | 0.32 |
| 6 | 5 | 0.73 | 0.65 | 0.85 | 0.73 | 0.31 |
| Mean | | 0.67 | 0.58 | 0.80 | 0.67 | 0.29 |

Five-fold cross-validation were performed on the MASS and two independent experts. Table III presents the fold partition with subject IDs and evaluation metrics on the MASS dataset. The proposed approach achieved a mean precision, recall, F1-score, average precision (AP), and IoU of 0.67, 0.61, 0.76, 0.67, and 0.28. Table III summarizes the fold partition and model performance on the DREAMS dataset. The mean values of the evaluation metrics were 0.67, 0.58, 0.80, 0.67, and 0.29. The observed stability of F1-scores across expert annotations indicates the robustness and consistency of the proposed approach.

### C. Estimation of Detected Temporal Overlaps

To further evaluate the temporal consistency between detected events and expert annotations, we quantified the overlap at the event level. Violin plots in Figure 6 compare the average durations detected by the model to the ground truth across the five folds, showing that the model predictions align closely with the actual durations on the MASS and DREAMS datasets. Figure 7 illustrates the joint distribution of onset and duration discrepancies for each fold. The scatter plots show a high concentration of data points, with both onset and duration discrepancies symmetrically distributed between [-2, 2] and [-1, 1]. The probability density curves in Figure 8 reveal a similar distribution across all five folds, indicating consistent patterns in the prediction results. Each fold, represented by a distinct color, exhibits a near-identical shape with a peak around zero, suggesting little variance between the folds. This uniformity implies that the data have a consistent error distribution across all folds in the cross-validation process. No significant differences were observed between the folds by Welch t-test.

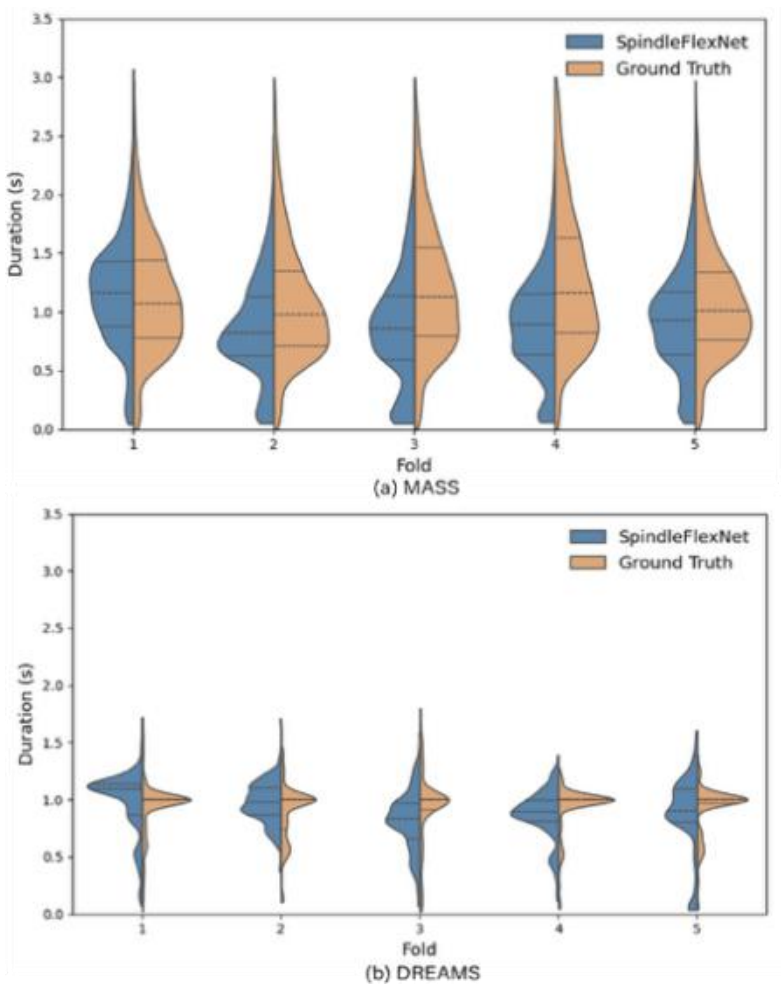

Fig. 6. Differences in spindle duration between the model predictions and ground truth in five-fold cross-validation on the MASS (top) and DREAMS (bottom) datasets.

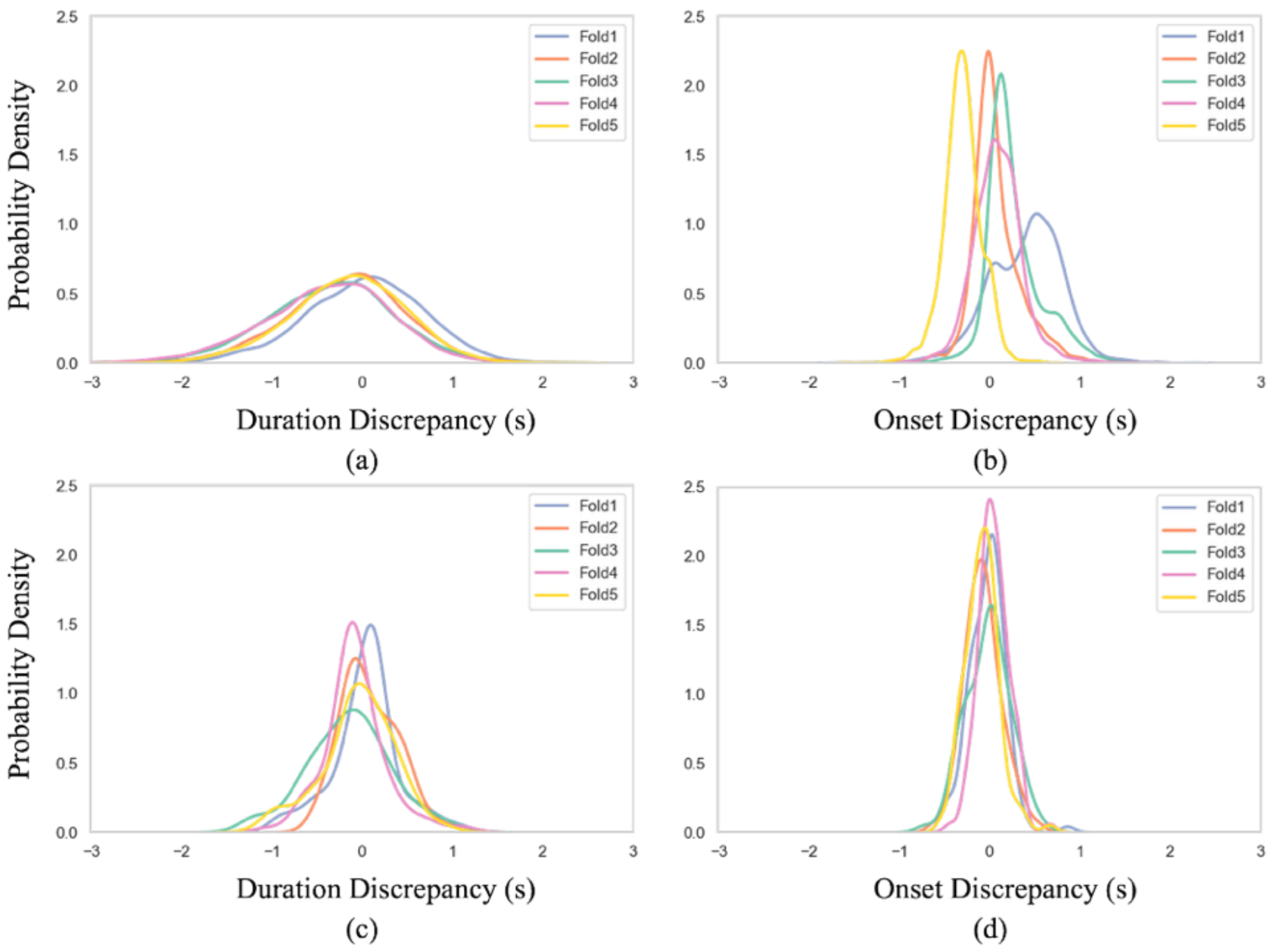

Fig. 7. Probability density curves of duration discrepancies across five folds on the MASS and DREAMS datasets.

TABLE IV

TIME EFFICIENCY COMPARISON BETWEEN HUMAN EXPERTS AND THE PROPOSED METHOD.

| methods | Number | Time spent (s) | Average time (s) | \|z\| | p |
|---|---|---|---|---|---|
| Human experts | 100 | 230 | 2.30 | 12.16 | $1.32\times10^{-38}$ |
| Our method | | 1.46 | 0.015 | | |

## D. Comparison of Time Efficiency

The clinical utility of the proposed model was evaluated by comparing its efficiency against human experts in terms of time required for spindle identification (Table IV). With 100 spindles as a reference benchmark, manual annotation by a trained sleep expert required an average of 2.30 seconds per spindle. In contrast, the deep learning model completed the same task in 0.015 seconds per spindle on average—less than one-fifth of the manual time. Statistical analysis using the Wilcoxon signed-rank test confirmed the significance of this reduction ($|z| = 12.16$, $P < 0.05$), demonstrating the potential of the model to streamline clinical workflows.

## E. Performance Comparison with Other Approaches

To evaluate the performance of the proposed method, we compared its performance with some approaches reported in the same databases. The performances of representative studies on the MASS dataset, including Schizophrenia Spindle [50] WT + Rule-based [51], Topographic Manual [52] and SST + RUSBoost [38], are summarized in Tables V. The comparisons with WT + Rule-based [51], Topographic Manual [52], A7 [53], SST + RUSBoost [38], Teager + bagging [39], SUMO [54] , and OpenSpindleNet [55] are summarized in Table VI. The results indicate that the proposed model achieves competitive performance for both the MASS and DREAMS datasets.

# IV. DISCUSION

In this study, we propose a novel object-detection-inspired deep learning framework, i.e. SpindleFlexNet, for sleep spindles detection by one-dimensional leverage of RetinaNet-based structure and anchor mechanism. To the best of our knowledge, it is the first exploration and application of a classical deep learning-based object detection algorithm in this field. This framework reflects a methodological transfer innovation, as it adapts principles from two-dimensional object detection into the temporal EEG domain for the first time. SpindleFlexNet learns feature representations and decision boundaries directly from data in an end-to-end manner, enabling adaptation to complex and variable spindle patterns across subjects. On the MASS dataset, our model achieved a precision of 0.76, a recall of 0.61, and an F1-score of 0.67. Similarly, on the smaller DREAMS dataset, the proposed framework maintained high robustness, reaching 0.67 F1-score and demonstrating consistent generalizability across subjects. These results highlight the advantage of end-to-end feature learning in spindle detection compared with traditional two-stage machine learning pipelines.

Sleep spindles are closely linked to memory consolidation, neural plasticity, and the diagnosis of neurological disorders. Accurate spindle detection is thus crucial for both basic neuroscience research and clinical evaluation. One-dimensional detection of spindles remains challenging due to their transient, non-stationary nature and the low separability from background oscillatory activity. Our model integrates assembled anchor

TABLE V

Performance Comparison with Representative Detection Approaches on the MASS Dataset with UNION annotations.

| Year | Method | Recall | Precision | F1-score |
|---|---|---|---|---|
| 2012 | Schizophrenia Spindle [1] [50] | 0.15 | 0.91 | 0.26 |
| 2012 | WT + Rule-based[1] [51] | 0.84 | 0.53 | 0.65 |
| 2013 | Topographic Manual[1] [52] | 0.46 | 0.89 | 0.61 |
| 2020 | SST+RUSBoost[1] [38] | 0.74 | 0.72 | 0.72 |
| 2025 | SpindleFlex-Net | 0.61 | 0.76 | 0.67 |

[1]These results are referenced from paper [39].

TABLE VI

Performance Comparison with Representative Detection Approaches on the DREAMS Dataset with UNION annotations.

| Year of publication | Method | Recall | Precision | F1-score |
|---|---|---|---|---|
| 2012 | WT + Rule-based[1] [51] | 0.85 | 0.28 | 0.42 |
| 2013 | Topographic Manual[1] [52] | 0.60 | 0.73 | 0.66 |
| 2019 | A7[2] [53] | 0.52 | 0.71 | 0.60 |
| 2020 | SST+RUSBoost [38] | 0.72 | 0.56 | 0.61 |
| 2021 | Teager+bagging [39] | 0.63 | 0.76 | 0.69 |
| 2022 | SUMO[2] [43] | 0.45 | 0.72 | 0.55 |
| 2025 | OpenSpindleNet[2] [55] | 0.73 | 0.68 | 0.69 |
| 2025 | SpindleFlex-Net | 0.58 | 0.80 | 0.67 |

[1]These results are referenced from paper [39]. [2]These results are referenced from paper [55].

generation, matching, and regression by utilizing a customized 1D IoU, 1D Focal Loss, and 1D NMS. This design enables dense and point-wise detection of multiple sleep spindles (1–7) within each 15-second EEG segment, demonstrating strong temporal precision and high computational efficiency. The implementation of 1D convolution and tailored loss function allows flexible and accurate training for onset and offset coordinate localization. Furthermore, the dense 1D anchor strategy ensures that closely occurring or partially overlapping spindles can be effectively separated, a challenge that has been insufficiently addressed in existing literature. These contributions highlight the potential application value of classical two-dimensional object detection algorithms for detecting high-difficulty and frequency-specific EEG waveforms.

The proposed method was primarily compared with approaches based on the union of two experts on the MASS and DREMS datasets. Experimental results demonstrate that, relative to traditional methods based on empirical rules as well as general machine learning methods, our approach achieves noticeable advantages in partial evaluation metrics. By leveraging the adaptive one-dimensional detection strategies, our framework shows a tendency toward improved precision while exhibiting relatively stable performance across different EEG segments. While individual differences in spindle and annotation characteristics may lead to fold-specific fluctuations, the overall five-fold results and cross-dataset results indicate consistent generalization performance. These observations indicate that RetinaNet-based method can serve as a promising alternative to empirical methods.

In subsequent analyses of onsets and durations between the MASS and DREAMS datasets, the onset and duration discrepancies between the reference events and the detected events are clustered across the five folds. Furthermore, the probability density curves of duration discrepancies of all folds are centered near zero, indicating that the proposed model achieves small prediction errors in most cases. Moreover, the similarity of the curves across folds demonstrates the robustness and stability of the method. This consistency across folds demonstrates the reliability and reproducibility of the method for participant-specific EEG segments. Meantime, the achieved averaged IoUs on the MASS and DREAMS datasets were 0.28 and 0.29, which are generally larger than the threshold of 0.20 in previous research [38, 40, 56]. These findings highlight the robustness of SpindleFlexNet in both temporal localization and duration estimation, underscoring its potential as a reliable approach for event detection in complex time-series data.

This study has some limitations. From a data perspective, the EEG recordings used in this study were sourced from healthy subjects in the MASS and DREAMS datasets. Further investigations could focus on training and evaluating the model from individuals with sleep-related diseases [57] or infant populations [19, 58], thereby broadening the clinical relevance and generalizability of the proposed approach. In future work, incorporating channel-specific physical features, such as frequency-domain characteristics of EEG signals will enhance model interpretability [59], thus partially addressing the black-box problem inherent in model development [60, 61]. Additionally, the study is based on supervised learning, incorporating unsupervised or weakly supervised learning methods may help reduce the dependence on manual annotations [62-65].

## V. CONCLUSION

In conclusion, we proposed SpindleFlexNet, a novel and flexible deep learning framework for accurate detection of sleep spindles in EEG signals. The framework incorporates a customized implementation of one-dimensional RetinaNet-based architecture and captures low-saliency events in EEG signals. Our approach enables dense anchor-based temporal event detectionof multiple spindles within EEG segments, while maintaining strong temporal sensitivity and high detection efficiency. This work demonstrates the feasibility and potential of extending object detection paradigms to sleep biomarker analysis on EEG signals. In future work, we aim to enhance the generalizability of the framework across diverse EEG datasets, explore real-time deployment in clinical settings, and extend the model to jointly detect other sleep micro-events such as K-complexes or high-frequency oscillations. Further integration with neurophysiological priors will be included to improve the robustness and clinical relevance of the proposed method.